\documentclass[epj]{svjour}
\usepackage{graphics}
\usepackage[misc]{ifsym}
\usepackage{extarrows}
\usepackage{epstopdf}
\usepackage[numbers,sort&compress]{natbib}
\usepackage{hyperref}
\hypersetup
{
hypertex=true,
colorlinks=true,
linkcolor=blue,
filecolor=blue,
urlcolor=blue,
citecolor=blue,
}
\begin{document}
\title{Electric Current Induced by Microwave Stark Effect of Electrons on Liquid Helium}
\author{T. Wang\inst{1} \and M. Zhang\inst{1} \thanks{\emph{e-mail:} zhangmiao@swjtu.edu.cn (corresponding author)}
\and L. F. Wei\inst{2} \thanks{\emph{e-mail:} lfwei@swjtu.edu.cn}
}
%
%
\institute{School of Physics Science and Technology, Southwest Jiaotong University, Chengdu 610031, China
\and School of Information Science and Technology, Southwest Jiaotong University, Chengdu 610031, China}
\date{Received: date / Revised version: date}
%
\abstract{
We propose a frequency-mixed effect of Terahertz (THz) and Gigahertz (GHz) electromagnetic waves in the cryogenic system of electrons floating on liquid helium surface. The THz wave is near-resonant with the transition frequency between the lowest two levels of surface state electrons. The GHz wave does not excite the transitions but generates a GHz-varying Stark effect with the symmetry-breaking eigenstates of electrons on liquid helium. We show an effective coupling between the inputting THz and GHz waves, which appears at the critical point that the detuning between electrons and THz wave is equal to the frequency of GHz wave. By this coupling, the THz and GHz waves cooperatively excite electrons and generate the low-frequency ac currents along the perpendicular
direction of liquid helium surface to be experimentally detected by the image-charge approach [Phys. Rev. Lett. {\bf 123}, 086801 (2019)]. This offers an alternative approach for THz detections.
} 
\maketitle
\section{Introduction}
\label{sec:1}

Electrons floating on liquid helium surface have been considered as an alternative physical system to implement quantum computation~\cite{Platzman1967,PhysRevX.6.011031,
PhysRevLett.89.245301,PhysRevB.67.155402,
PhysRevLett.107.266803,PhysRevLett.101.220501,
PhysRevLett.105.040503,Monarkha2019,
PhysRevLett.124.126803,Spin,PhysRevB.86.205408}. Sommer first noticed that there is a barrier at the surface of liquid helium that prevents electrons from entering its interior~\cite{PhysRevLett.12.271}. Therefore, in addition to the attractive force of image charge in liquid helium, the surface electron forms a one-dimensional (1D) hydrogen atom above liquid helium surface~\cite{RevModPhys.46.451,PhysRevLett.32.280,PhysRevB.13.140}.
The two lowest levels of the 1D hydrogen atoms have been proposed to encode
the qubits, which have good scalability due to the
strong Coulomb interactions between electrons~\cite{Platzman1967}. However, the single electron detection is still very difficult in experiments. Recently, E. Kawakami et.al. have used the image-charge approach to experimentally detect the Rydberg states of surface electrons on liquid helium~\cite{PhysRevLett.123.086801}. Compared with the conventional microwave absorption method, this method may be more effective to detect the quantum state of surface electrons~\cite{Elarabi2021,PhysRevLett.126.106802}.

The transition frequency between the lowest two levels of surface electrons is in the regime of Terahertz (THz) and therefore has perhaps applications in the fields of THz detection and generation. The THz spectrum is between the usual microwave and infrared regions, and has many practical applications~\cite{NaturePhotonics}, such as THz wireless communication~\cite{THZ-Wireless Communication}, astronomical observation~\cite{EPJP-astronomy}, Earth's atmosphere monitoring~\cite{watter}, and the food safety~\cite{food}. Specially, in biological and medical fields~\cite{biology,APL}, THz imaging have smaller damages in body than X-ray, but higher resolution than mechanical supersonic waves.

In this paper, we show a frequency-mixed effect of the THz and the Gigahertz (GHz) electromagnetic waves in the system of electrons floating on liquid helium. This effect could be applied to detect THz radiation. In our mixer, the inputting THz wave near-resonantly excites the transition between the lowest two levels of surface state electrons. The GHz wave does not excite the transitions but generates a GHz-varying Stark effect. This Stark effect has no counterpart in the system of natural atoms. The effective coupling between natural atoms and microwaves needs either the atomic high Rydberg states of $n\approx47$~\cite{Jing2020} or the magnetic coupling (with the strong microwave inputting). Here, the surface-bound states of electrons are symmetry broken due to the barrier at the liquid surface, and therefore the GHz wave can cause a GHz-varying energy gap of electrons on liquid helium by the lowest order electric dipole interaction. Specially, we find an effective coupling between the inputting THz and GHz waves, which appears at the frequencies condition that the detuning between electrons and THz wave is equal to the frequency of GHz wave. By this coupling,
the THz and GHz waves cooperatively excite electrons and generate the low-frequency ac currents along the perpendicular direction of liquid helium surface to be experimentally detected by using the image-charge approach~\cite{PhysRevLett.123.086801,Elarabi2021,PhysRevLett.126.106802}.

\section{The Hamiltonian}
\label{sec:2}

Due to the image potential and the barrier at liquid helium
surface, electronic motion along the perpendicular direction of liquid helium surface can be treated as a 1D hydrogen
atom. Its energy level is well known as ~\cite{PhysRevA.61.034901,PhysRevA.80.055801}
\begin{equation}
\begin{aligned}
E_n=-\frac{\Lambda^{2}e^{4}m_e}{2n^{2}\hbar^{2}}\,,
\end{aligned}
\label{E1}
\end{equation}
with $\Lambda=(\varepsilon-1)/4(\varepsilon+1)$. The
$e$ is the charge of an electron, $m_e$ is the electronic mass, and $\varepsilon= 1.0568$ the dielectric constant of liquid helium. Numerically, the transition frequency between the lowest two levels is $\omega_{12}=(E_2-E_1)/\hbar\simeq0.1~$THz (corresponding to a wavelength of $\lambda\simeq3$~mm).
The eigenfunction of the above energy level reads
\begin{equation}
\begin{aligned}
\psi_{n}(z)=2n^{-5/2}r_{B}^{-3/2}z \exp\left(\frac{-z}{nr_B}\right)L_{n-1}^{(1)}\left(\frac{2z}{nr_{B}}\right)\,,
\end{aligned}
\label{E2}
\end{equation}
with $r_{B}=\hbar^{2}/(m_{e}e^{2}\Lambda)\simeq76\rm{\AA}$ being the Bohr radius, and
\begin{equation}
\begin{aligned}
L_{n}^{(\alpha)}(x)=\frac{e^{x}x^{-\alpha}}{n!}\frac{{\rm d}^{n}}{{\rm d}x^{n}}
(e^{-x}x^{n+\alpha})\,,
\end{aligned}
\label{E3}
\end{equation}
the Laguerre polynomial.

Considering only the lowest two levels in Eq.~(\ref{E1}), the Hamiltonian of an electron driven by the classical electromagnetic fields can be written as
\begin{equation}
\begin{aligned}
\hat{H}_0=\hat{H}_{00}+\hat{V}\,,
\end{aligned}
\label{E4}
\end{equation}
with the bare-atomic part
\begin{equation}
\begin{aligned}
\hat{H}_{00}=E_1|1\rangle\langle1|+E_2|2\rangle\langle2|\,,
\end{aligned}
\label{E5}
\end{equation}
and the electric dipole interaction
\begin{equation}
\begin{aligned}
\hat V&=e\textbf{z}\cdot \textbf{E} =(\textbf{u}_{11}\hat{\sigma}_{11}+\textbf{u}_{22}\hat{\sigma}_{22}
+\textbf{u}_{12}\hat{\sigma}_{12}+\textbf{u}_{21}\hat{\sigma}_{21})\cdot\textbf{E}\,.
\end{aligned}
\label{E6}
\end{equation}
Here, $\textbf{u}_{ij}=e\langle i|\textbf{z}|j\rangle=e\textbf{z}_{ij}$ is the electric dipole moment, and $\hat{\sigma}_{ij}=|i\rangle\langle j|$ the two-level operator with $i,j=1,2$. Note that $\langle 1|\textbf{z}|1\rangle \neq 0$ and $\langle 2|\textbf{z}|2\rangle \neq 0$ for the wave function (\ref{E2}). On electric field $\textbf{E}$, we first consider the case of two-wave mixing, i.e.,
\begin{equation}
\begin{aligned}
\textbf{E}(t)&=\textbf{E}_{T}\cos(\omega_{T}t+\theta_{T})
+\textbf{E}_{G}\cos(\omega_{G}t+\theta_{G})\,,
\end{aligned}
\label{E7}
\end{equation}
with $\omega_{T}\simeq 0.1~$ THz and $\omega_{G}\simeq 1~$GHz. The $\textbf{E}_{T}$ and $\textbf{E}_{G}$ are the amplitudes of the inputting THz and GHz waves. The $\theta_{T}$ and $\theta_{G}$ are their initial phases, respectively.

Limiting within the usual rotating-wave approximation, the $\omega_{G}$ works only on the first two terms in Eq.~(\ref{E6}), and $\omega_{T}$ works only on the last two terms in Eq.~(\ref{E6}).
Then, the Hamiltonian (\ref{E4}) in the usual interacting picture [defined by the unitary operator $\hat{U}_0=\exp(-i\hat{H}_{00}t/\hbar)$] can be approximately written as
\begin{equation}
\begin{aligned}
\hat{H}=\hbar\left[\Omega_{11}(t)
\hat{\sigma}_{11}+\Omega_{22}(t)\hat{\sigma}_{22}
+\Omega_{12}(t)\hat{\sigma}_{12}+\Omega_{21}(t)\hat{\sigma}_{21}\right]\,,
\end{aligned}
\label{E8}
\end{equation}
with the time-dependent coupling frequencies $\Omega_{11}(t)=2\Omega_{110}
\cos(\omega_{G}t+\theta_{G})$ ,
$\Omega_{22}(t)=2\Omega_{220}\cos(\omega_{G}t+\theta_{G})$, and
$\Omega_{12}(t)=\Omega_{21}^{\ast}(t)=\Omega_{120}\exp(it\Delta)\exp(i\theta_{T})$. The
$\Omega_{110}=\textbf{u}_{11}\cdot\textbf{E}_{G}/(2\hbar)$ and
$\Omega_{220}=\textbf{u}_{22}\cdot\textbf{E}_{G}/(2\hbar)$ are due to the symmetry-breaking eigenstates of electron floating on liquid helium surface. The
$\Omega_{120}=\textbf{u}_{12}\cdot\textbf{E}_{T}/(2\hbar)$ is the standard Rabi frequency of electric dipole transition, and $\Delta=\omega_T-\omega_{12}$ is the detuning between the THz wave and the resonant frequency $\omega_{12}$ of electronic transition $|1\rangle\rightleftharpoons|2\rangle$.
For the analyzable effects of THz and GHz mixing, we apply another unitary transformation $\hat{U}=\exp[-i\hat{h}(t)/\hbar]$ to the electronic states. Here,
\begin{equation}
\begin{aligned}
\hat{h}(t)
=\hbar[\theta_{11}(t)\hat{\sigma}_{11}+\theta_{22}(t)\hat{\sigma}_{22}]\,,
\end{aligned}
\label{E9}
\end{equation}
\begin{equation}
\begin{aligned}
\theta_{11}(t)=\int_{0}^{t}\Omega_{11}(t){\rm d}t
=\frac{2\Omega_{110}}{\omega_{G}}\sin(\omega_{G}t+\theta_{G})\,,
\end{aligned}
\label{E10}
\end{equation}
and
\begin{equation}
\begin{aligned}
\theta_{22}(t)=\int_{0}^{t}\Omega_{22}(t){\rm d}t
=\frac{2\Omega_{220}}{\omega_{G}}\sin(\omega_{G}t+\theta_{G})\,.
\end{aligned}
\label{E11}
\end{equation}
As a consequence, Hamiltonian (\ref{E8}) in the new interacting picture reads
\begin{equation}
\begin{aligned}
\hat{H}_{I}=\hbar\Omega_{120}
\left[e^{i\phi(t)}\hat{\sigma}_{12}+
e^{-i\phi(t)}\hat{\sigma}_{21}\right]\,,
\end{aligned}
\label{E12}
\end{equation}
with
\begin{equation}
\begin{aligned}
e^{i\phi(t)}=e^{i\theta_T}
e^{i\Delta t}e^{i\xi\sin(\omega_{G}t+\theta_{G})}\,,
\end{aligned}
\label{E13}
\end{equation}
and
\begin{equation}
\begin{aligned}
\xi=\frac{2(\Omega_{110}-\Omega_{220})}{\omega_{G}}\,.
\end{aligned}
\label{E14}
\end{equation}

According to the well-known Jacobi-Anger expansion, we approximately write (\ref{E13}) as
\begin{equation}
\begin{aligned}
e^{i\phi(t)}=&e^{i\theta_T}
e^{i\Delta t}\sum\limits_{n=-\infty}^{\infty}
J_{n}(\xi)e^{in(\omega_{G}t+\theta_{G})}\\
\\
=&J_0(\xi)e^{i\theta_T}e^{i\Delta t}
+J_{-1}(\xi)
e^{i[(\Delta-\omega_G)t+(\theta_T-\theta_G)]}
+J_{1}(\xi)
e^{i[(\Delta+\omega_G)t+(\theta_T+\theta_G)]}\\
\\
&+\cdots\,,
\end{aligned}
\label{E15}
\end{equation}
with $J_{n}(\xi)=(-1)^nJ_{-n}(\xi)$ being the Bessel function of the first kind. Considering the field strength of GHz wave is weak, i.e., $\xi\ll1$, and supposing the frequency condition $\Delta=\omega_G\gg\Omega_{120}$ is satisfied, then the Eq.~(\ref{E15}) reduces to
\begin{equation}
\begin{aligned}
e^{i\phi(t)}\approx&J_{-1}(\xi)
e^{i(\theta_T-\theta_G)}\approx-\frac{\xi}{2}e^{i(\theta_T-\theta_G)}\,,
\end{aligned}
\label{E16}
\end{equation}
with the rotating-wave approximation. As a consequence, Hamiltonian (\ref{E12}) reduces to
\begin{equation}
\begin{aligned}
\hat{H}_{R}=\hbar\Omega_{\rm eff}
\left[e^{i(\theta_T-\theta_G)}\hat{\sigma}_{12}+
e^{-i(\theta_T-\theta_G)}\hat{\sigma}_{21}\right]\,,
\end{aligned}
\label{E17}
\end{equation}
which generates the standard Rabi oscillation at frequency $\Omega_{\rm eff}=\Omega_{120}(\Omega_{220}-\Omega_{110})/\omega_G$.
This Hamiltonian implies that the electrons floating on liquid helium may be applied as a frequency-mixer to generate the low frequency signal with $\Omega_{\rm eff}\ll \omega_G\ll \omega_T$.

\section{The master equation}
\label{sec:3}
We use the standard master equation of classical electromagnetic waves interacting with the two-levels system~\cite{RevModPhys.77.633} to numerically solve the dynamical evolution of surface state electrons on liquid helium. The master equation reads
\begin{equation}
\begin{aligned}
\frac{\rm{d}\hat{\rho}}{\rm{d}t}=-\frac{i}{\hbar}[\hat{H},\hat{\rho}]
+\hat{L}(\hat{\rho})\,,
\end{aligned}
\label{E18}
\end{equation}
with the decoherence operator
\begin{equation}
\begin{aligned}
\hat{L}(\hat{\rho})=&\frac{\Gamma_{21}}{2}
(2\hat{\sigma}_{12}\hat{\rho}\hat{\sigma}_{21}-\hat{\sigma}_{22}\hat{\rho}-\hat{\rho}\hat{\sigma}_{22})\\
&+\frac{\gamma_{1}}{2}(2\hat{\sigma}_{11}\hat{\rho}\hat{\sigma}_{11}-\hat{\sigma}_{11}\hat{\rho}-\hat{\rho}\hat{\sigma}_{11})\\
&+\frac{\gamma_{2}}{2}(2\hat{\sigma}_{22}\hat{\rho}\hat{\sigma}_{22}-\hat{\sigma}_{22}\hat{\rho}-\hat{\rho}\hat{\sigma}_{22})\,.
\end{aligned}
\label{E19}
\end{equation}
Above, $\hat{\rho}=|\psi\rangle\langle\psi|=\sum_{i,j}^2\rho_{ij}|i\rangle\langle j|$, with $\sum_{i=1}^2\rho_{ii}=1$ and $\rho_{ij}=\rho_{ji}^{\ast}$. The $\rho_{ij}$ is the so-called density matrix elements of the two-levels system. The $\Gamma_{21}$ is the spontaneous decay rate of $|2\rangle\rightarrow|1\rangle$. The $\gamma_{1}$ and $\gamma_{2}$ describe the energy-conserved dephasing~\cite{RevModPhys.77.633}.
It is worth to note that the unitary operator $\hat{U}=\exp[-i\hat{h}(t)/\hbar]$ made for the Hamiltonian transformation (\ref{E8}) $\rightarrow$ (\ref{E12}) does not change the formation of master equation. The proofs are detailed  as follows.

Firstly, we rewrite the density operator as
$\hat{\rho}=|\psi\rangle\langle\psi|
=\hat{U}\hat{\rho}_{I}\hat{U}^{\dagger}$
with $\hat{\rho}_{I}=|\psi'\rangle\langle\psi'|$, and rewrite the Hamiltonian (\ref{E8}) as $\hat{H}=\hat{H}_{I0}+\hat{H}_{\rm int}$, with $\hat{H}_{I0}=\hbar\left[\Omega_{11}(t)
\hat{\sigma}_{11}+\Omega_{22}(t)\hat{\sigma}_{22}\right]$ and
$\hat{H}_{\rm int}=\hbar\left[\Omega_{12}(t)\hat{\sigma}_{12}
+\Omega_{21}(t)\hat{\sigma}_{21}\right]$. Our unitary operator
obeys the commutation relation $[\hat{U},\hat{H}_{I0}]=0$, so
\begin{equation}
\begin{aligned}
\frac{d\hat{\rho}}{dt}&=\frac{d\hat{U}}{dt}\hat{\rho}_{I}
\hat{U}^{\dagger}+\hat{U}\frac{d\hat{\rho}_{I}}{dt}\hat{U}^{\dagger}
+\hat{U}\hat{\rho}_{I}\frac{d\hat{U}^{\dagger}}{dt}\\
&=\hat{U}\frac{d\hat{\rho}_{I}}{dt}\hat{U}^{\dagger}
+\frac{-i}{\hbar}[\hat{H}_{I0},\hat{U}\hat{\rho}_{I}\hat{U}^{\dagger}]\,.
\end{aligned}
\label{E20}
\end{equation}
Secondly, according to the original master equation (\ref{E18}), we have
\begin{equation}
\begin{aligned}
\frac{d\hat{\rho}}{dt}&=\frac{-i}{\hbar}[\hat{H},\hat{\rho}]+\hat{L}(\hat{\rho})\\
&=\frac{-i}{\hbar}
[\hat{H}_{\rm int},\hat{U}\hat{\rho}_{I}\hat{U}^{\dagger}]
+\frac{-i}{\hbar}[\hat{H}_{I0},\hat{U}\hat{\rho}_{I}\hat{U}^{\dagger}]
+\hat{L}(\hat{\rho})\\
&=\frac{-i}{\hbar}\hat{U}[\hat{H}_{I},
\hat{\rho}_{I}]\hat{U}^{\dagger}
+\frac{-i}{\hbar}[\hat{H}_{I0},\hat{U}\hat{\rho}_{I}\hat{U}^{\dagger}]
+\hat{L}(\hat{\rho})\,,
\end{aligned}
\label{E21}
\end{equation}
and where $\hat{H}_{I}=\hat{U}^\dagger\hat{H}_{\rm int}\hat{U}$ is nothing but Eq.~(\ref{E12}).
The decoherence operator obeys the following equation
\begin{equation}
\begin{aligned}
\hat{L}(\hat{\rho})&=\hat{U}\hat{L}(\hat{\rho}_{I})\hat{U}^{\dagger}\,,
\end{aligned}
\label{E22}
\end{equation}
because $\hat{U}^{\dagger}\hat{\sigma}_{11}\hat{U}=\hat{\sigma}_{11}$, $\hat{U}^{\dagger}\hat{\sigma}_{22}\hat{U}=\hat{\sigma}_{22}$, and
$\hat{\sigma}_{12}\hat{\rho}\hat{\sigma}_{21}
=\hat{U}\hat{\sigma}_{12}\hat{\rho}_{I}\hat{\sigma}_{21}\hat{U}^\dagger$,
with $\hat{U}^{\dagger}\hat{\sigma}_{12}\hat{U}
=\hat{\sigma}_{12}\exp[i\xi\sin(\omega_{G}t+\theta_{G})]$ and $\hat{U}^{\dagger}\hat{\sigma}_{21}\hat{U}
=\hat{\sigma}_{21}\exp[-i\xi\sin(\omega_{G}t+\theta_{G})]$.

Finally, comparing Eqs. (\ref{E20}) and (\ref{E21}), we find
\begin{equation}
\begin{aligned}
\hat{U}\frac{d\hat{\rho}_{I}}{dt}\hat{U}^{\dagger}
=\frac{-i}{\hbar}\hat{U}[\hat{H}_{I},
\hat{\rho}_{I}]\hat{U}^{\dagger}
+\hat{U}\hat{L}(\hat{\rho}_{I})\hat{U}^{\dagger}\,,
\end{aligned}
\label{E23}
\end{equation}
and get the master equation in interaction picture
\begin{equation}
\begin{aligned}
\frac{d\hat{\rho}_{I}}{dt}
=\frac{-i}{\hbar}[\hat{H}_{I},\hat{\rho}_{I}]
+\hat{L}(\hat{\rho}_{I})\,,
\end{aligned}
\label{E24}
\end{equation}
which has the same form as that of (\ref{E18}). Moreover, using the above approach one can easily find that the usual unitary transformation $\hat{U}_0$ also does not change the form of master equation.
Compared with the original Hamiltonian (\ref{E8}), the Hamiltonian $H_I$ in interacting pictures is more clear to show the frequency-matching condition between electron and microwaves. While, the Hamiltonian $H_I$, as proved above, is also valid to numerically (exactly) solve the master equation.

\section{The results and discussion}
\label{sec:4}
According to master equation (\ref{E24}), we get the following equations for density matrix elements,
\begin{align}
\label{E25}
\frac{{\rm d}\rho_{22}}{{\rm d}t} &
=i\rho_{21}\Omega_{120}e^{i\phi(t)}
-i\rho_{12}\Omega_{120}e^{-i\phi(t)}-\Gamma_{21}\rho_{22}\,,\\
\label{E26}
\frac{{\rm d}\rho_{12}}{{\rm d}t}&=
i(1-2\rho_{22})\Omega_{120}e^{i\phi(t)}
-\frac{1}{2}(\Gamma_{21}+\gamma_{1}+\gamma_{2})\rho_{12}\,.
\end{align}
In the experimental systems~\cite{PhysRevLett.123.086801,Elarabi2021,PhysRevLett.126.106802}, the liquid helium surface (with the surface-state electrons) is set
approximately midway between two plates of a parallel-plate
capacitor, and the induced image charges on one of the capacitor plates is described by $Q_{\rm image}\approx
Q_{e}\langle z\rangle/D$.
The $Q_{e}=eN$ is the charge of $N$ electrons on liquid helium surface,  and $\langle z\rangle$ the average height of these electrons. The $D$ is the distance between the two parallel plates of capacitor.
In terms of density matrix elements, the expectation value $\langle z\rangle$ in Schr\"{o}dinger picture is described by
\begin{equation}
\begin{aligned}
\langle z\rangle
=&\rho_{11}z_{11}+\rho_{22}z_{22}+2 z_{21}{\rm Re}\left[\rho_{12}e^{i\omega_{12}t}
e^{-i\xi\sin(\omega_{G}t+\theta_{G})}\right]\\
=&z_{11}+(z_{22}-z_{11})\rho_{22}\\
&+2 z_{21}{\rm Re}\left[{\rm Re}(\rho_{12})e^{i\omega_{12}t}
e^{-i\xi\sin(\omega_{G}t+\theta_{G})}
+{\rm Im}(\rho_{12})e^{i\omega_{12}t}
e^{-i\xi\sin(\omega_{G}t+\theta_{G})}e^{i\frac{\pi}{2}}\right]\,.
\end{aligned}
\label{E27}
\end{equation}
Numerically, the transition matrix elements are $z_{11}=1.5r_{B}$, $z_{22}=6r_{B}$, and $z_{21}=-0.5587r_{B}$.
Note that, the term $\exp(i\omega_{12}t)$ in Eq.~(\ref{E27}) is due to the standard unitary transformation $\hat{U}_0$ in Sec. \ref{sec:2} applied for the interacting Hamiltonian (\ref{E8}), and $\exp[-i\xi\sin(\omega_{G}t+\theta_{G})]$ is due to the second unitary transformation $\hat{U}$ employed for Hamiltonian (\ref{E12}). Specially, the term $\exp(i\omega_{12}t)=\exp[i(\omega_{T}-\Delta)t]$ rapidly vibrates with the intrinsic frequencies of two-level electrons on liquid helium and generates the THz radiation.

\begin{figure}[tpb]
\centering
\resizebox{1\textwidth}{!}{\includegraphics{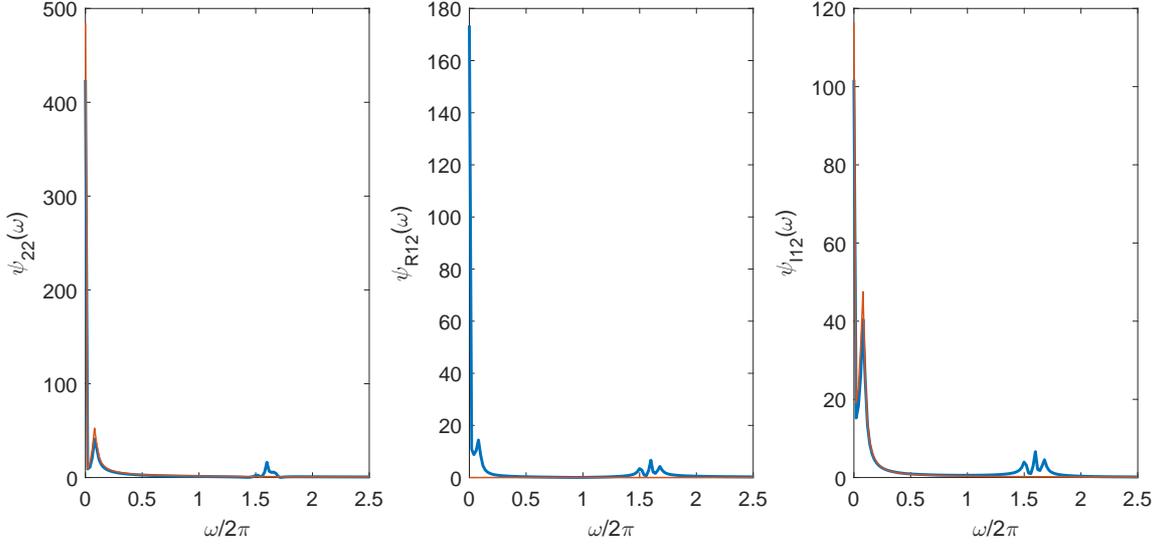}}
\caption{The numerical solutions of frequency distribution functions $\psi_{22}(\omega)$, $\psi_{R12}(\omega)$, and $\psi_{I12}(\omega)$, based on Eqs.~(\ref{E25}) and (\ref{E26}) of  density matrix elements. The values of the necessary parameters in this figure are $\xi=0.5$, $\Omega_{120}=0.1\omega_{G}$, $\Delta=\omega_G=1$~GHz, $\Gamma_{21}=10$~MHz, and $\gamma_1=\gamma_2=0.5\Gamma_{21}$.
The red lines are obtained from the rotating-wave approximated Hamiltonian (\ref{E17}), and which can be simply explained by the analytical solution $\rho_{ij}$ of  Rabi oscillation.
The blue lines are obtained from the Hamiltonian (\ref{E12}), without the rotating-wave approximation. The blue lines of $\psi_{R12}(\omega)$ and $\psi_{I12}(\omega)$ show that the vibrational frequencies of $\rho_{12}$ are far smaller than THz and thus the off-diagonal term in Eq.~(\ref{E27}) is relate to the THz vibration. The low frequency signal in $\psi_{22}(\omega)$ can be amplified effectively by the low-noise amplifier~\cite{Elarabi2021}, while the high frequency components in $\psi_{22}(\omega)$ could be filtered out by the low-pass filter in experiments.}
\label{fig:1}
\end{figure}

According to our rotating-wave approximated Hamiltonian (\ref{E17}), the density matrix elements $\rho_{22}$ and $\rho_{12}$ vibrate with the Rabi frequency $2\Omega_{\rm eff}\ll\Omega_G\ll\Omega_T$. The rotating-wave terms in Eq.~(\ref{E15}), such as $\exp(i\Delta t)$, vibrate with much higher frequencies than $\Omega_{\rm eff}$, but has small amplitude as shown in Fig.~\ref{fig:1}. The frequency distribution functions in Fig.~\ref{fig:1} are obtained by numerically solving the following Fourier transforms
\begin{equation}
\begin{aligned}
\rho_{22}(T)=\int_{-\infty}^{\infty}\psi_{22}(\omega)e^{-i\omega T}{\rm d}\omega\,,
\end{aligned}
\label{E28}
\end{equation}
\begin{equation}
\begin{aligned}
{\rm Re}(\rho_{12}(T))=\int_{-\infty}^{\infty}\psi_{R12}(\omega)e^{-i\omega T}{\rm d}\omega\,,
\end{aligned}
\label{E29}
\end{equation}
and
\begin{equation}
\begin{aligned}
{\rm Im}(\rho_{12}(T))=\int_{-\infty}^{\infty}\psi_{I12}(\omega)e^{-i\omega T}{\rm d}\omega\,.
\end{aligned}
\label{E30}
\end{equation}
The $\psi_{22}(\omega)$ is the amplitude of a harmonic vibrational component in function $\rho_{22}(T)$, with $T=\Omega_{120}t$ being the dimensionless time. The frequency of this component (harmonic vibration) is $\omega \Omega_{120}$, with $\omega$ being the dimensionless coefficient. The explanations for $\psi_{R12}(\omega)$ and $\psi_{I12}(\omega)$ are similar.

\begin{figure}[htpb]
\centering
\resizebox{0.7\textwidth}{!}{\includegraphics{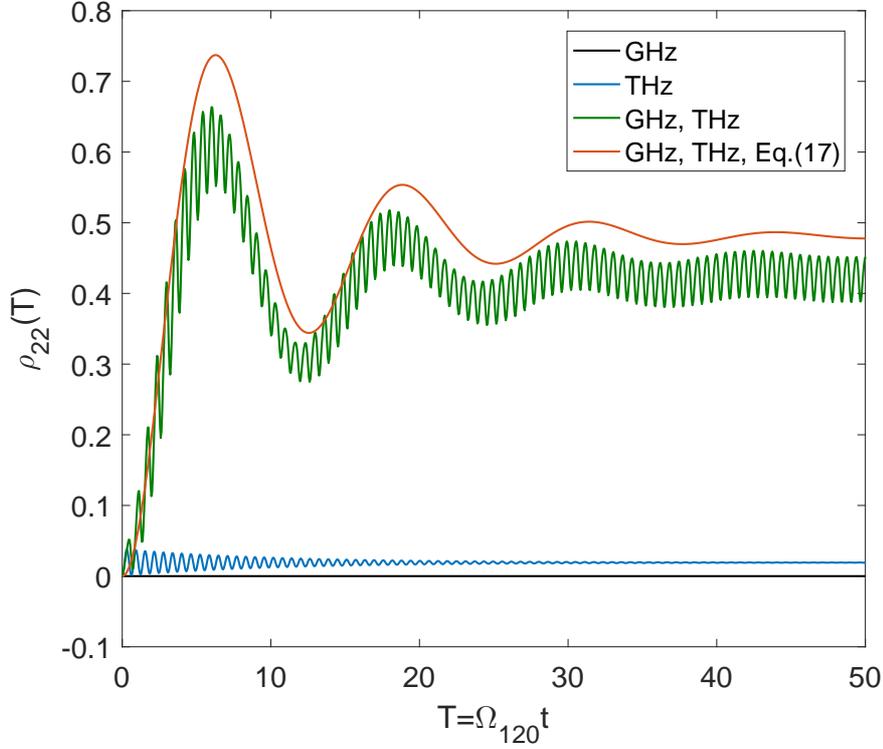}}
\caption{Numerical solutions of $\rho_{22}$ for the case of two-wave mixing.
The black line (closing
to the horizontal axis) describes the driving by GHz wave only, with $\xi=0.5$, $\omega_G=1$~GHz, and $\Omega_{120}=0$, the electron remains at the ground state, and does not generate any signals. The blue line, with $\xi=0$, $\Omega_{120}=0.1\omega_{G}$, and $\Delta=\omega_G=1$~GHz, describes the excitation by THz wave only. The green line, with $\xi=0.5$, $\Omega_{120}=0.1\omega_{G}$, and $\Delta=\omega_G=1$~GHz, describes the cooperative excitation by two waves, the $\rho_{22}$ becomes bigger and has the component of lower frequency vibration, $\Omega_{\rm eff}\ll \omega_G\ll \omega_T$. The values of decay and dephasing are same as that used in Fig.~\ref{fig:1}. Similarly, the red line here is the solution from the rotating-wave approximated Hamiltonian~(\ref{E17}), and the other lines are obtained from the Hamiltonian~(\ref{E12}) without the rotating-wave approximation.}
\label{fig:2}
\end{figure}

Due to the negligible components of high frequency vibrations in $\rho_{12}$, see Fig.~\ref{fig:1}, the off-diagonal term in Eq.~(\ref{E27}) is still the THz vibration due to the characteristic term $\exp(i\omega_{12} t)$. The THz waves travel in free space, and thus the off-diagonal term in Eq.~(\ref{E23}) is negligible for the  microwave circuit experiments~\cite{PhysRevLett.123.086801,Elarabi2021,PhysRevLett.126.106802}, and then the formula of the vibrational image charge reduces to
\begin{equation}
\begin{aligned}
Q_{\rm image}(t)&=
\frac{Q_{e}(z_{22}-z_{11})}{D}\rho_{22}\,.
\end{aligned}
\label{E31}
\end{equation}
As to was mentioned at the beginning of this paper, the GHz wave can not directly excite the transition $|1\rangle\rightleftharpoons|2\rangle$, the THz wave acts as a trigger to start the system work. This mechanism is showed in Fig.\ref{fig:2}, where the detuning between THz wave and electron is set at $\Delta=\omega_G$. The electric field strength and the frequency of the GHz microwave are set at $E_{G0}=1$~V/cm and $\omega_G=1$~GHz, respectively.

In Fig.~\ref{fig:2}, the red line is the solution of the rotating-wave approximated Hamiltonian $\hat{H}_{R}$, i.e., Eq.~(\ref{E17}). This low frequency signal is the damped one because the Hamiltonian $\hat{H}_{R}$ is time-independent. For a time-independent Hamiltonian, the master equation has the steady-state solution, i.e., ${\rm d}\rho_{ij}/{\rm d}t=0$ with $t\rightarrow\infty$~\cite{ZHANG201412}. The same phenomenon can be found in the blue line in Fig.~\ref{fig:2}, which describes the excitation by THz wave only. In this case, the Hamiltonian (\ref{E12}) reduces to $\hat{H}_{\rm blue}=\hbar\Omega_{120}
[\exp(i\Delta t+i\theta_T)\hat{\sigma}_{12}+
\exp(-i\Delta t-i\theta_T)\hat{\sigma}_{21}]$ with $\xi=0$. This Hamiltonian can be also written in a time-independent form, i.e.,  $\hat{H}'_{\rm blue}=\hbar\Omega_{120}
[\exp(i\theta_T)\hat{\sigma}_{12}+
\exp(-i\theta_T)\hat{\sigma}_{21}]-\hbar\Delta |1\rangle\langle 1|$, by simply employing the unitary transformation of $\exp(i\Delta t|1\rangle\langle 1|)$.
So, the blue line is also the damped one. Readers may notice that the blue line vibrates much faster than the red line. This can be simply explained by the analytic solution of Schr\"{o}dinger equation of $\hat{H}'_{\rm blue}$. The solution says the electron excitation probability $P_e=[\Omega_{120}^2/(2\Delta^2_{\Omega})]\times[1-\cos(2\Delta_{\Omega}t)]$
with the frequency $\Delta_{\Omega}=\sqrt{(\Delta^2/4)+\Omega_{120}^2}$\,.  In the large detuning regime, i.e., $\Delta\gg\Omega_{120}$, the vibrational frequency $\Delta_{\Omega}$ is obviously larger than the standard Rabi frequency $\Omega_{120}$, but the amplitude is small (the rotating-wave effect).

The green line is the numerical solution of Hamiltonian (\ref{E12}) without the rotating-wave approximation. Then, the vibration has not only the low frequency component but also the high frequency component. In this case, the persistent oscillation exists, because Hamiltonian (\ref{E12}) is associated with the rotating-wave terms and can not be written in the time-independent form by any unitary transformation, as well as the original Hamiltonian (\ref{E8}). However, the frequency of persistent oscillation is on the order of the inputting GHz wave and is not the usable signal that a frequency-mixer should produce. Thus we suggest using the low frequency signal predicted by the red line to detect the THz inputting, although the signal is the damped one. This means that a pulsed source of GHz microwave is needed. Furthermore, we note that the above detuning is set at $\Delta=\omega_G$. This condition may not exactly satisfy in the practical experiments, and therefore the effective Hamiltonian (\ref{E17}) becomes
\begin{equation}
\begin{aligned}
\hat{H}_{R}=\hbar\Omega_{\rm eff}
\left[e^{i(\theta_T-\theta_G)}e^{i\delta t}\hat{\sigma}_{12}+
e^{-i\delta t}e^{-i(\theta_T-\theta_G)}\hat{\sigma}_{21}\right]\,,
\end{aligned}
\label{E32}
\end{equation}
with $\delta=\Delta-\omega_G$. The solution of this Hamiltonian is similar to that of $\hat{H}_{\rm blue}$, but can still generate the low-frequency signal with $\sqrt{(\delta^2/4)+\Omega_{\rm eff}^2}$ and $\delta\sim\Omega_{\rm eff}$.

\begin{figure}[tpb]
\centering
\resizebox{0.7\textwidth}{!}{\includegraphics{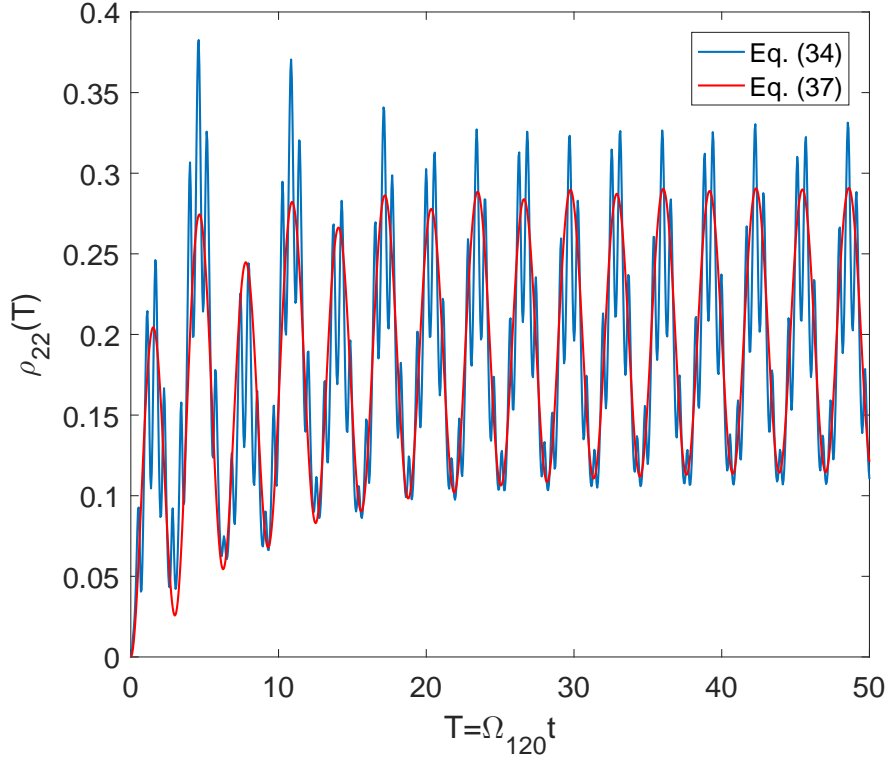}}
\caption{Numerical solutions of $\rho_{22}$ for the three-wave mixing (\ref{E34}), with $\Delta=1$~GHz, $\omega_G=0.9$~GHz, $\omega_{G2}=1.1$~GHz, $\xi=\xi_2=0.5$,  $\theta_G=\theta_{G2}$, $\Omega_{120}=0.1\Delta$, $\gamma_1=\gamma_2=0.5\Gamma_{21}$, and $\Gamma_{21}=10$~MHz. The blue and red lines are obtained from Hamiltonian (\ref{E34}) and its rotating-wave approximated version (\ref{E37}), respectively.}
\label{fig:3}
\end{figure}

Finally, we describe a possibility to generate the persistent signal, i.e., monitor THz incoming online. Based on the above discussion, we apply another GHz wave $\textbf{E}_{G2}\cos(\omega_{G2}t+\theta_{G2})$ to the electrons. Such that the inputting field (\ref{E7}) extends to
\begin{equation}
\begin{aligned}
\textbf{E}'(t)&=\textbf{E}_{T}\cos(\omega_{T}t+\theta_{T})
+\textbf{E}_{G}\cos(\omega_{G}t+\theta_{G})
+\textbf{E}_{G2}\cos(\omega_{G2}t+\theta_{G2})\,,
\end{aligned}
\label{E33}
\end{equation}
and consequently, the Eqs.~(\ref{E12}) and (\ref{E13}) become, respectively,
\begin{equation}
\begin{aligned}
\hat{H}'_{I}=\hbar\Omega_{120}
\left[e^{i\phi'(t)}\hat{\sigma}_{12}+
e^{-i\phi'(t)}\hat{\sigma}_{21}\right]\,,
\end{aligned}
\label{E34}
\end{equation}
and
\begin{equation}
\begin{aligned}
e^{i\phi'(t)}=e^{i\theta_T}
e^{i\Delta t}e^{i\xi\sin(\omega_{G}t+\theta_{G})}
e^{i\xi_2\sin(\omega_{G2}t+\theta_{G2})}\,.
\end{aligned}
\label{E35}
\end{equation}
The derivation for $\xi_2=2[\Omega_{110}(\textbf{E}_{G2})-\Omega_{220}(\textbf{E}_{G2})]/\omega_{G2}$ is similar to that of parameter $\xi$ in Eq.~(\ref{E14}). Considering also $\xi_2\ll1$, we have
\begin{equation}
\begin{aligned}
e^{i\phi'(t)}\approx-\frac{\xi}{2}
e^{i(\Delta-\omega_G)t}e^{i\theta_{TG}}
-\frac{\xi_2}{2}e^{i(\Delta-\omega_{G2})t}e^{i\theta_{TG2}}\,,
\end{aligned}
\label{E36}
\end{equation}
by employing the rotating-wave approximation again, and where $\theta_{TG}=\theta_{T}-\theta_{G}$ and $\theta_{TG2}=\theta_{T}-\theta_{G2}$.
Furthermore, considering the typical case that $\xi\approx\xi_2$ and $\delta=\Delta-\omega_G=\omega_{G2}-\Delta$, such that the effective Hamiltonian (\ref{E17}) becomes,
\begin{equation}
\begin{aligned}
\hat{H}'_{R}\approx\hbar\Omega_{\rm eff}
\left\{e^{i\theta_{TG}}\left[e^{i\delta t}+e^{-i\delta t}e^{i(\theta_{G}-\theta_{G2})}\right]\hat{\sigma}_{12}+
e^{-i\theta_{TG}}\left[e^{-i\delta t}+e^{i\delta t}e^{-i(\theta_{G}-\theta_{G2})}\right]\hat{\sigma}_{21}\right\}\,.
\end{aligned}
\label{E37}
\end{equation}
Due to the small detuning $\delta$, this Hamiltonian can be never written in a time-independent form by any unitary transformations, and thereby the master equation has no steady-state solution. This causes the persistently oscillating charge, as shown in Fig.~\ref{fig:3}, which serves as a source for the low-frequency microwave outputting and could be applied to realize the real-time detection of THz incoming.

\section{Conclusion}
In summary, we have studied the frequency-mixed effects of THz and GHz waves in the cryogenic system of electrons floating on liquid helium. Different from the natural atoms, the transition frequency between the lowest two levels of electrons floating on liquid is in the THz regime. Moreover,  the surface-state of electrons is symmetry breaking due to the barrier at the liquid surface. Therefore, both of THz and GHz waves can effectively drive the electrons via the electric dipole interaction. Specifically, the THz wave near-resonantly excites the transition between the lowest two levels of surface-bound electrons, the GHz wave does
not excite the transition but generates the GHz-varying Stark effect. Using the unitary transformation approach and the rotating-wave approximation, we found an effective Hamiltonian (\ref{E17}) of two-wave mixing with the detuning $\Delta=\omega_{12}-\omega_T=\omega_G$, which could generate the significant ac current of frequency much lower than GHz, as shown in Fig.~\ref{fig:2}. The Hamiltonian is time-independent, so the generated low-frequency signal is the damped one due to the decay of surface-state electrons. To generate the persistent low-frequency signal, the time-dependent Hamiltonian is required, for example, the effective Hamiltonian (\ref{E37}) proposed for the case of three-wave mixing.

\textbf{Acknowledgements}:
This work was supported by the National Natural Science Foundation of China, Grants No. 12047576, and No. 11974290.

\textbf{Data Availability Statement}:
The data generated or analyzed during this study are included in this published article.

\end{document}